\documentclass[prl,aps,twocolumn,superscriptaddress%
]{revtex4-2}

\usepackage{amsmath}
\usepackage{graphicx}
\usepackage{dcolumn}
\usepackage{bm}

\usepackage{graphicx}
\usepackage{dcolumn}
\usepackage{bm}
\usepackage{tikz}
\usepackage{amssymb}   
\usetikzlibrary{shapes} 

\definecolor{tabblue}{rgb}{0.1216, 0.4667, 0.7059}
\definecolor{taborange}{rgb}{1.0000, 0.4980, 0.0549}
\definecolor{tabgreen}{rgb}{0.1725, 0.6275, 0.1725}
\definecolor{tabred}{rgb}{0.8392, 0.1529, 0.1569}
\definecolor{tabpurple}{rgb}{0.5804, 0.4039, 0.7412}
\definecolor{tabbrown}{rgb}{0.5490, 0.3373, 0.2941}
\definecolor{tabpink}{rgb}{0.8902, 0.4667, 0.7608}
\definecolor{tabgray}{rgb}{0.4980, 0.4980, 0.4980}
\definecolor{tabolive}{rgb}{0.7373, 0.7412, 0.1333}
\definecolor{tabcyan}{rgb}{0.0902, 0.7451, 0.8118}

\begin{document}

\preprint{APS/123-QED}

\title{Long distance interaction between particles in a soap film}

\author{Youna Louyer}
\affiliation{Univ. Rennes, CNRS, IPR – UMR 6251. Rennes, France}
\author{Megan Delens}
\affiliation{GRASP, Institute of Physics B5a, University of Liège, B4000 Liège, Belgium.}
\author{Nicolas Vandewalle}
\affiliation{GRASP, Institute of Physics B5a, University of Liège, B4000 Liège, Belgium.}
\author{Benjamin Dollet}
\affiliation{Université Grenoble Alpes, CNRS, LIPhy, 38000 Grenoble France}
\author{Isabelle Cantat}
\affiliation{Univ. Rennes, CNRS, IPR – UMR 6251. Rennes, France}
\author{Ana\"is Gauthier}
\affiliation{Univ. Rennes, CNRS, IPR – UMR 6251. Rennes, France}
\email{anais.gauthier@univ-rennes.fr}




\date{\today}

\begin{abstract}

Millimeter-sized particles trapped at the surface of a liquid bath attract each other through the deformation of the liquid-air interface, a phenomenon known as ``the Cheerios effect''.  We consider here a  situation  similar  at  first  sight:   the  interaction  between  two millimeter-sized particles trapped in an horizontal soap film.
In this geometry, the deformation of the film due to the weight of one particle extends over the entire system size, which induces an extremely long-ranged attraction. Combined with the low viscous friction in the film, this leads to intricate particle orbits, lasting up to ten seconds before the two particles eventually collide. 

By tracking the particles dynamics, we measure the force exerted by each particle on the other, and we develop a theoretical model. Because the interface deformation induced by a particle depends on its position in the soap film, the attractive force has two features that fundamentally depart from classical interaction forces. The force exerted by one particle on the other differs both in direction and magnitude from the reverse interaction, with an asymmetry reaching 150\% when one particle is close to the center and the other one close to the frame. Reciprocity is recovered when both particles are close to the film center. These results are a original example of non-reciprocal effective interactions due to boundary conditions.

\end{abstract}


\maketitle


\section{\label{Introduction}Introduction} 

Particles trapped at the surface of a liquid bath rarely remain at rest. They spontaneously move towards each other and eventually aggregate \cite{nicolson1949, vella2005}. Capillary attraction makes fluid interfaces a powerful tool for driving the motion of particles (or in nature, insects or seeds) \cite{loudet2005, peruzzo2013, Delens2025} and to guide their self-assembly \cite{bowden1997, gart2011, liu2018, ko2022, hooshanginejad2024}, enabling the engineering of new two-dimensional, potentially reconfigurable materials \cite{pieranski1980, irvine2010, aubry2008, vandewalle2020}. Capillary attraction arises from the deformation of the liquid interface induced by the wetting properties of the particles and by their weight. The theoretical form of the interaction force is now well established \cite{kralchevsky2001, vella2005, dominguez2008} and recent experiments have quantified it at the micrometric \cite{carrasco2019} and millimetric \cite{ho2019, delens2023} scales. When particles deform the interface by their weight, the characteristic range of the force is set by the capillary length: as a result, capillary forces are 
short-ranged for millimeter-sized objects, restricting the interaction to particles separated by distances smaller than their own size. 

While the interaction between particles trapped at liquid interfaces has been investigated extensively, much less is known about free-standing liquid films, where the particles -- with a diameter much larger than the film thickness -- bridge two interfaces. Soap films, in particular, are known to effectively capture small solids or liquid drops \cite{legoff2008, gilet2009, stogin2018}. Once captured, the particles (or droplets) display a variety of unusual dynamics, including long-lasting oscillations \cite{louyer2025}, orbiting motion \cite{martischang2026} or spontaneous spatial ordering -- where partially wet particles assemble into lines \cite{shi2024}. These dynamics result from two distinct effects. First, particles locally increase the film thickness around them, generating a short-ranged attraction between neighboring menisci \cite{dileonardo2008, shi2024}. In addition, their weight induces a macroscopic deformation of the film, which gives rise to an attraction mediated by the film itself \cite{martischang2026}. 

\smallskip

Here, we focus on the film-mediated force on one particle induced by the presence of the other one, which we call the "interaction force". More precisely, noting 1 and 2 the two particles in the film, we call $F_{1 \rightarrow 2}$ the force applying on particle 1 and due to particle 2, and $F_{1 \rightarrow 2}$ the force on particle 2 due to particle 1. We show that the interaction force is extremely long-ranged, allowing particles to interact over distances of the order of the film size. This leads to complex trajectories, with particles circling each other for tens of seconds. We provide the first experimental measurement of this force, using two complementary approaches: by analyzing particle dynamics, and via the magnetic actuation of paramagnetic particles. In contrast with particles trapped at a liquid bath, the interaction force loses its symmetry in a soap film (with $F_{2 \rightarrow 1} \neq F_{1 \rightarrow 2}$) and is not necessarily oriented along the interparticle axis. Our experiments capture this asymmetry and quantitatively match the proposed model. 

\section{\label{Experiment}Orbiting motion}

\begin{figure}
\begin{center}
\includegraphics[width=0.99\columnwidth]{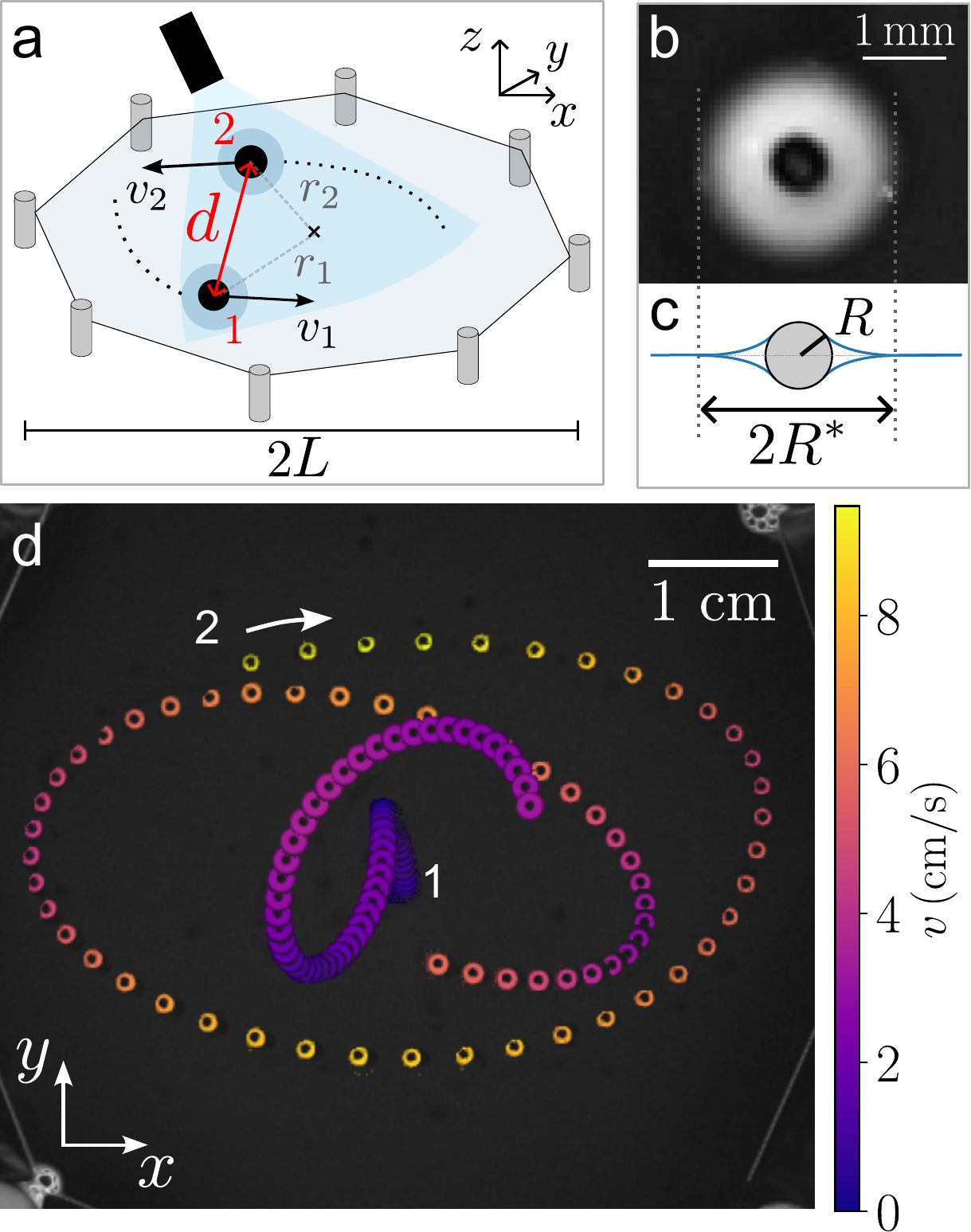}
\caption{\textbf{a.} Experimental setup. Two particles are deposited on an horizontal soap film, and their motion is recorded from the top. \textbf{b.} Top view picture of a particle with radius $R$ = 0.5 mm trapped in a soap film, obtained by fluorescence imaging. The bead is surrounded by a meniscus, visible as a brighter ring of radius $R^*$. \textbf{c.} Exact shape of the meniscus of size $R^*$ (seen from the side) calculated  using Ref. \cite{orr1975}. \textbf{d.} Colored chronophotography showing the successive positions (separated by $\Delta t=20$ ms) of particles 1 and 2 as they circle each other. The color code gives the particles velocity, varying from 0 cm/s (dark blue) up to 9 cm/s (yellow). The position of particles 1 and 2 at $t = 0$ is respectively noted by the numbers 1 and 2.
}
\label{Figure1}
\end{center}
\end{figure}

In an experiment, two spherical particles, with radius $R$ (250~$\mu$m~$< R < $ 750~$\mu$m), density $\rho$ (2580 kg/m$^3$ $<  \rho <$ 9200 kg/m$^3$) and mass $m$ are successively deposited into a horizontal soap film using tweezers. As shown in Fig. \ref{Figure1}a, the film is supported by a 
nylon wire stretched between eight vertical pillars, forming an octagonal frame of effective diameter $2L = 7.4$ cm (between two opposite sides). The position of the particles with respect to the center of the film is noted $\mathbf{r_1} = (x_1,y_1)$ for particle 1 and $\mathbf{r_2} = (x_2,y_2)$ for particle 2. The particle's velocities are respectively $\bm{v_1}$ and $\bm{v_2}$, and the distance separating them is $d$. In all experiments, the film thickness is constant and equal to $e =$ 8 $\pm$ 1~$\mu$m, which is obtained by withdrawing the frame from a soap solution at a controlled speed of 3 mm/s with a motorised translation stage. 
The soap solution has a surface tension $\gamma = 33.2 \pm 0.1$ mN/m and a density $\rho_f =$ 1042 kg/m$^3$. It is made of a solution of sodium dodecyl sulfate (SDS) at 5.6 g/L (2.4 times the critical micelle concentration) and dodecanol (50 mg/L) in a water-glycerol mixture containing $15\%_{vol}$ of glycerol. In addition, 0.8 g/L of fluorescein is added to visualize local variations of the film thickness.

Once deposited, the pre-wetted particles remain trapped in the film, provided they are sufficiently small and light \cite{louyer2025}. Their diameter is two orders of magnitude larger than the film thickness $e$, so that they protrude significantly on both sides of the film. Each particle is thus surrounded by a liquid meniscus, which grows with time as the liquid from the film is drawn towards it by capillary suction \cite{aradian2001, guo2019}. Experimentally, the meniscus is visualized using fluorescence imaging, and appears as a bright ring of radius $R_{\rm m}$ around the particle (Figure \ref{Figure1}b). The corresponding meniscus shape (calculated using Ref. \cite{orr1975}) is shown in side view in Figure \ref{Figure1}c. In the experiments reported here, the particles are sufficiently far apart that their menisci do not overlap.

\smallskip

Particle 1 is deposited first, and moves towards the center of the frame, a spontaneous motion due to the parabolic deformation of the film under its own weight \cite{louyer2025}. After approximately $10$ seconds, at a time $t = 0$, particle 2 is introduced in the film, typically 3 cm apart from particle 1, and gently pushed in the orthoradial direction. The particle dynamics are recorded from the top, using a high-speed camera (Phantom Miro LAB3a10). Figure \ref{Figure1}c evidences the two trajectories for $t > 0$; the color code indicates the particles velocities, which range between 0 cm/s (purple) and 9 cm/s (yellow). Despite the large distance $d \gg R$ between the particles, especially at small time, the attraction of particle 2 is able to significantly move particle 1 away from its equilibrium position at the film center. The trajectories that follow form a pattern which reflects the competition between two forces: attraction towards the film center and a mutual attraction between the particles. Due to the very small friction in a soap film (which arises primarily from a viscous shear stress in air \cite{louyer2025}) this orbiting motion can last more than ten seconds before the two menisci surrounding the particles eventually touch (see also Supplementary Movie 1). 

\smallskip

The aim of this paper is to characterize and model the long-ranged force at the origin of the particle's orbits.

\smallskip

\section{\label{Results}Measurement of the interaction force}

 The interaction force is first measured dynamically, using the variations of the particle's momentum with time. This method requires a previous knowledge of the other forces acting on the particle, in particular the drag force \cite{vella2005, vassileva2005, dalbe2011, gauthier2019}. Here, these forces can be measured independently at the start of each experiment, before the second particle is deposited. 
 
More specifically, the equation of motion for particle 1 writes:
 
 \begin{equation}
m_1^*  \frac{d\mathbf{v_1}}{dt}= \bm{F}_{\rm film} + \bm{F}_{\rm \eta}
\label{dynamics}
\end{equation}
 
 where $\bm{F}_{\rm film}$ is the total force acting on particle 1 and mediated by the film, and $\bm{F}_{\rm \eta}$ is a drag force, damping the motion of particle 1. Here, $m_1^*$ is the mass of the moving object, \textit{i.e.}, the system consisting of particle 1 and its meniscus, which we refer to as "particle 1" for simplicity. As shown in Supplementary Movie 1, the meniscus moves almost as a rigid body together with the particle -- which is a consequence of the two-dimensional nature of the flow in a soap film \cite{louyer2025}. The mass $m_1^*$ increases slowly over time (see Supplementary Figure 1), by less than 6\% in the duration of an entire experiment. For measurement-related reasons, its instantaneous value is used when possible; otherwise an average value is used (for $t<0$).
 
 \medskip

\begin{figure*}
\begin{center}
\includegraphics[width=\linewidth]{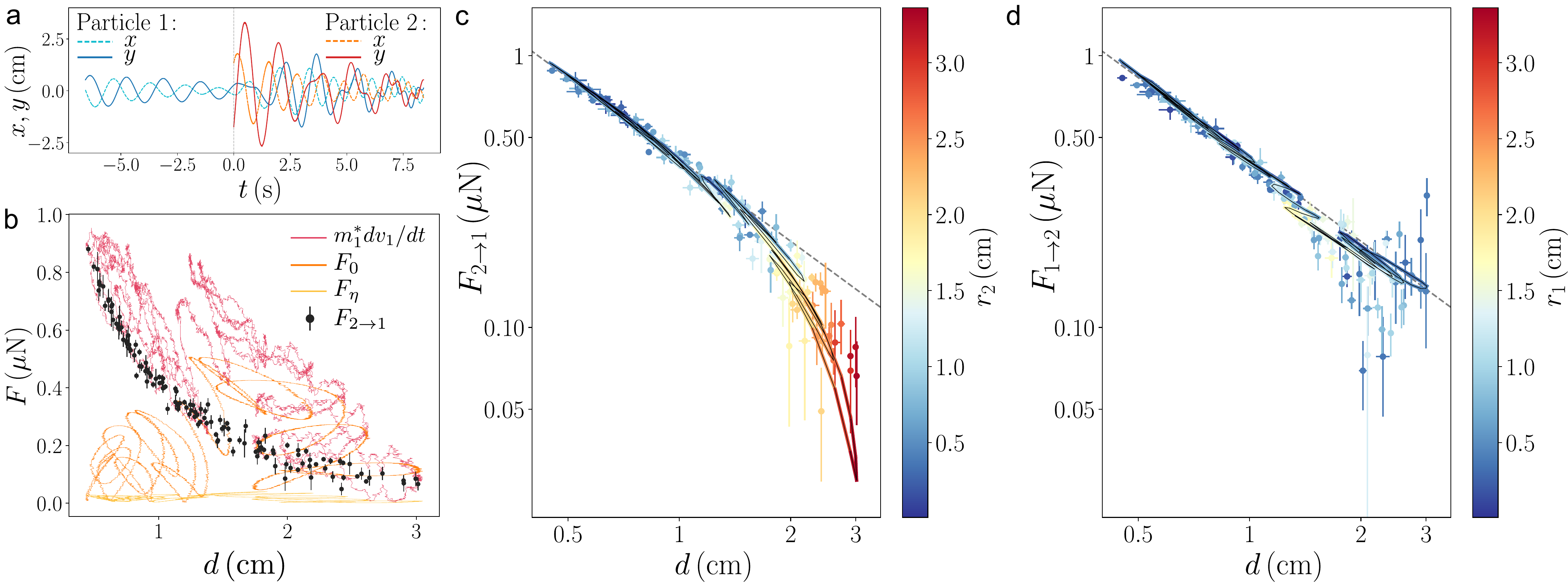}
\caption{\textbf{a}. Position $(x_1, y_1)$ of particle 1 (light and dark blue) and  $(x_2, y_2)$ of particle 2 (orange and red) as a function of time $t$, for the same experiment as in Figure \ref{Figure1}c. For $t<0$, particle 1, alone in the film, slides in damped harmonic oscillations, allowing the measurement of the force $\bm{F}_{0}$ driving its motion and the friction force $\bm{F}_{\eta}$ ($m_1^*$ is here taken as the average mass for $t<0$). At $t = 0$ particle 2 is introduced into the film. \textbf{b}. Force balance on particle 1. Inertia $m_1^* dv_1/dt$ (red line), $F_0$ (orange line) and friction $F_{\eta}$ (yellow) are calculated for each time step and plotted as a function of the inter-particle distance $d$. The interaction force $F_{2\rightarrow 1}$ is deduced from Newton's second law (equation \ref{dynamics}), using the exact measurement of $m_1^*(t)$ at each time. Inertia, $F_0$ and $F_{\eta}$ do not depend directly on $d$ and therefore form a scatter plot. However, the force $F_{2 \to 1}$ (black dots), deduced at each moment from these elements, varies smoothly with $d$. The error bars correspond to the standard deviation over 100 ms. \textbf{c}. Force $F_{2 \rightarrow 1}$ as a function of the distance $d$ between the two particles. Each point is the time average and standard deviation of the force during 100 ms. The color indicates the distance $r_2$ between particle 2 and the center of the film. \textbf{d}. Force $F_{1 \rightarrow 2}$ as a function of $d$, the color code indicating here $r_1$. In \textbf{c.} and \textbf{d.}, the continuous black line is the full theoretical prediction (equation \ref{Force_bipolar}). The dotted gray line is equation \ref{force_symmetric}.} 
\label{Figure2}
\end{center}
\end{figure*}

 We consider first the motion of particle 1 alone in the film, for $t<0$.  As shown in Figure \ref{Figure2}a, particle 1 moves in damped harmonic oscillations towards the center of the film. This indicates that \textit{i)} the film force driving the particle towards the center is spring-like: $\bm{F}_{\rm film} = \mathbf{F}_{\rm 0} = -k_1\mathbf{r_1}$, and \textit{ii)} the drag force is viscous: $\mathbf{F}_{\eta} = -\alpha_1 \mathbf{v_1}$. These forces have been characterized in our earlier study \cite{louyer2025}: $\bm{F_0}$ arises from the deflection of the film under its own weight, and varies linearly with the film thickness $e$ and the mass $m$ of the moving object, while the friction force $\mathbf{F}_{\eta}$ depends on the Boussinesq number $Bo$ which compares dissipation in air and within the film. The coefficients $k_1$ and $\alpha_1$ are determined for each experiment at $t< 0$ from the oscillation period and the damping time of particle 1 as it slides into the film. 

 
 \smallskip
 
 \noindent When a second particle is introduced in the film at $t = 0$, the total force $\bm{F}_{\rm film}$ acting on particle 1 is modified by an additional contribution arising from the presence of particle 2. We define the interaction force $\bm{F}_{2 \rightarrow 1}$ as:
 
 \begin{equation}
 \bm{F}_{2 \rightarrow 1} = \bm{F}_{\rm film} - \bm{F}_0
 \end{equation}
 
\noindent \textit{i.e.} as the additional force that appears when the second particle is introduced into the film. This definition relies on the additivity of the forces (an assumption justified in the Supplementary Materials) and it yields $F_{2 \rightarrow 1}~\propto~m_1^* m_2^*$, as expected for an interaction force.

\medskip



In Figure \ref{Figure2}b, the four terms of equation \ref{dynamics} are plotted as a function of the interparticle distance $d$, for the same experiment as in Figure \ref{Figure2}a. The amplitude of the friction force $F_{\rm \eta}$ (yellow line) is typically one order of magnitude smaller that the other forces. The spring-like force $F_{0}$ (in orange) and the inertia (in red) vary with the distance $r_1$ of the particle to the center or with time: they are thus scattered when plotted as a function of $d$. However, the force $F_{2 \rightarrow 1}$ (black dots), deduced at each timestep from the other three measurements using equation \ref{dynamics}, decreases smoothly as a function of the interparticle distance $d$, as would be expected from an interaction force. Each data point shows the average and the standard deviation of $F_{2 \rightarrow 1}$ over a time interval of 100 ms. The same data is represented in log-log in Figure \ref{Figure2}c, the color code indicating the value of the distance $r_2$ of particle 2 with respect to the film center. The black line is the full theoretical model (Equation \ref{Force_bipolar}), without adjustable parameter. The gray dotted line (with slope -1 in log-log) is Equation \ref{force_symmetric}. For small $d$ and $r_2$, the force $F_{2 \rightarrow 1}$ decreases as $1/d$. For $r_2 > 1.5$ cm (which also corresponds to $d > 1.5$ cm) $F_{2 \rightarrow 1}$ deviates from this scaling and decreases faster than $1/d$. 

\smallskip

The force $F_{1 \rightarrow 2}$ is also measured dynamically using the same method as $F_{2 \rightarrow 1}$, but using here $k_2 = m_2^*/m_1^* k_1$ (which is a direct consequence of the proportionality of $k$ with the mass of the particle) and $\alpha_2 = \alpha_1$. $F_{1 \rightarrow 2}$ is plotted as a function of $d$ in Figure \ref{Figure2}d, with the color code now representing $r_1$. Each point shows the average and standard deviation of the force during 100 ms, the dotted line is equation \ref{force_symmetric} and the black line Equation \ref{Force_bipolar}. The experimental data is scattered at small times $t$ (corresponding the larger distances $d$), which we explain by vertical oscillations of particle 2 in the film in the first seconds after its deposition, that temporarily impact the horizontal force balance (equation \ref{dynamics}). Here, $r_1$ remains relatively small at all times, and the force $F_{2 \rightarrow 1}$ follows the $1/d$ scaling up to $d = 3$ cm. 

\smallskip

\noindent Strikingly, the comparison between $F_{1 \rightarrow 2}$ and $F_{2 \rightarrow 1}$ evidences an asymmetry in the interaction force: for $d \simeq 3$ cm, $F_{2 \rightarrow 1}$ is on average 1.5 times smaller than $F_{1 \rightarrow 2}$. This asymmetry is observed in different experiments (see Supplementary Figures 2 and 3), and happens when one particle is close to the center and the other close to the side of the film. We now discuss and model this phenomenon.

\bigskip

\section{\label{Model}Model}

To derive theoretically the force $\bm{F}_{2 \rightarrow 1}$, we assume the frame to be circular. The soap film is identified with a surface $S$ of equation $z = h(\bm{r})$, deformed by its own weight and by the weights $m_1^* g$ and $m_2^* g$ of the two particles that it holds. The film shape is given by the balance along the normal to each film element between the Laplace pressure $2 \gamma \kappa$ (with $\kappa$ the local film curvature) and the hydrostatic pressure $\rho_f g e \cos\theta$ (with $\rho_f$ the liquid density, $e$ the film thickness and $\theta$ the local angle between the vertical axis and the normal to the film) \cite{cohen2017, martischang2025, louyer2025}. In the limit of small slopes (here $\lVert \nabla^{2D} h \rVert \sim h/L \simeq 10^{-2} \ll 1$), the curvature can be linearized and the film profile $h$ obeys: 

\begin{equation}
\Delta h = \frac{\rho_f g e}{2 \gamma}. 
\label{film_eq}
\end{equation}

\noindent with $h = 0$ at the film frame and a vertical force balance at the perimeter $l_i$ of particle $i$ (with $i = \{1,2\}$), where the surface tension force balances the particle weight : 
\begin{equation}
- 2 \gamma \oint_{l_i} \bm{\nabla}h \cdot \bm{n}\,dl_i = m_i^* g,
\label{boundary_conditions}
\end{equation}
with $\bm{n}$ the outward normal, pointing towards the particle in the ($x,y$) plane.

\smallskip

Due to the linearity of equation \ref{film_eq} and the superposability of the boundary conditions, the film surface writes: $h(\bm{r}) = h_0(\bm{r}) + h_1(\bm{r; r_1}) + h_2(\bm{r; r_2})$, with $h_0$ the deformation of the film under its own weight, $h_1$ the deformation due to the weight of particle 1 alone at the position $\bm{r}_1$ in a weightless film, and $h_2$ the deformation due to particle 2 alone at the position $\bm{r}_2$. 

In this configuration, the exact derivation of the force $\bm{F}_{2 \rightarrow 1}$ is determined from the potential energy of the system \{film + particle 1 + particle 2\} (see Supplementary Materials). This yields an equivalent of the Nicolson's superposition approximation \cite{nicolson1949} for a soap film: $F_{2 \rightarrow 1}$ is exactly equal to the product of the weight of particle 1 with the 2D gradient of the interfacial displacement due to particle 2, estimated at the position of particle 1:

\begin{equation}
    \bm{F}_{2 \rightarrow 1} = - m_1^* g \frac{\partial h_2(\bm{r}_1;\bm{r}_2)}{\partial \bm{r}_1}
    \label{force}
\end{equation}

\noindent with $\partial/\partial \bm{r}_1$ the 2D gradient relative to the variable $\bm{r}_1$.

\begin{figure}
\begin{center}
\includegraphics[width=0.99\linewidth]{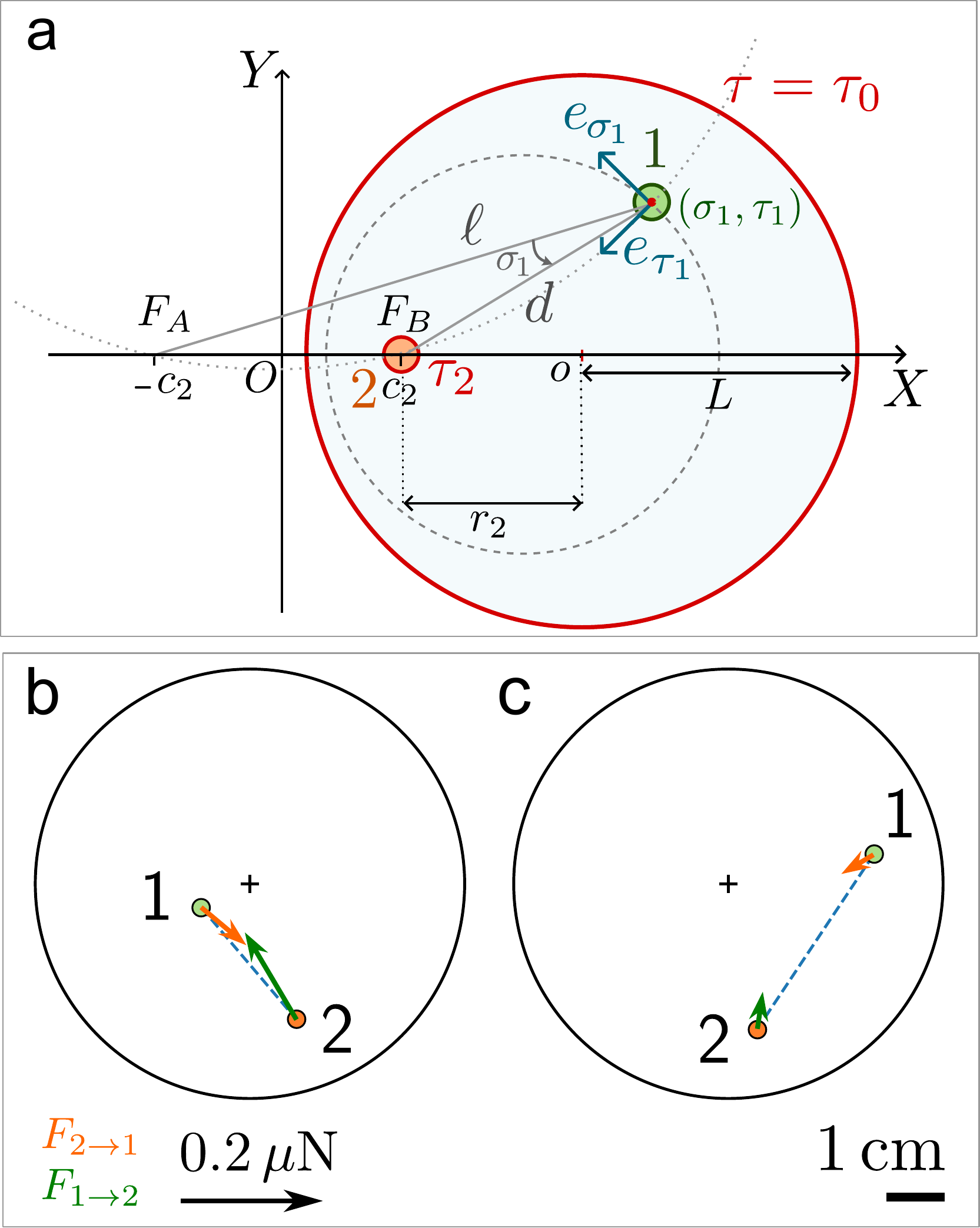}
\caption{\textbf{a.} In bipolar coordinates, the frame and the perimeter of particle 2 (in red) are iso-$\tau$ curves. Particle 2 is centered on one of the foci, in $X = +c_2$. Particle 1, placed at any position in the frame has coordinates $\sigma_1, \tau_1$. The force $\bm{F}_{2 \rightarrow 1}$ applying on particle 1 is oriented along the direction $\bm{e}_{\tau}$. \textbf{b-c.} Theoretical forces $\bm{F}_{2 \rightarrow 1}$ and $\bm{F}_{1 \rightarrow 2}$ (shown respectively with orange and green arrows) for two different particle positions in the film (\textbf{b} is a situation encountered experimentally.) The interaction force is asymmetric: the force amplitudes $F_{2 \rightarrow 1}$ and $F_{1 \rightarrow 2}$ can differ by a factor 1.5, and the forces are not colinear with the interparticle direction.} 
\label{Figure3}
\end{center}
\end{figure}

\smallskip
We now seek to express the deformation $h_2$, which is, the deformation induced by particle 2 alone in a weightless film. $h_2$ is solution to the Laplace equation $\Delta h_2 = 0$ with the boundary conditions: \textit{i)} $h_2 = 0$ at the film frame, and \textit{ii)} Eq. \ref{boundary_conditions} along the diameter of particle 2. Given the geometry of the problem, we solve $h_2$ using bipolar coordinates in a local Cartesian coordinate system ($X O Y$) attached to particle 2. The $X$-axis is defined as the line joining the center of particle 2 to the frame center $o$ (see Figure \ref{Figure3}a). Bipolar coordinates are based on two foci $F_A$ and $F_B$, located at $X = -c_2$ and $X = +c_2$, respectively. Any point $M$ of the ($X,Y$) plane has coordinates ($\sigma, \tau$) where $\sigma$ is the angle $F_AMF_B$ and $\tau = \ln(\ell/d)$ with $\ell$ (resp $d$) the distance between $F_A$ (resp. $F_B$) to $M$. With a right choice of $c_2$, the circular frame and the equator of particle 2 (shown in red in Figure \ref{Figure3}) are both expressed as iso-$\tau$ curves, respectively with coordinates $\tau = \tau_0$ and $\tau = \tau_2$. Solving $\Delta h_2 = 0$ in bipolar coordinates is then straightforward (see Supplementary Materials) and yields $h_2 = -\frac{m_2^* g}{4 \pi \gamma}(\tau-\tau_0)$. The deformation $h_2$ due to a single particle thus varies linearly with $\tau$, \textit{i.e} logarithmically with the distance to the particle. Using the expression of the gradient in bipolar coordinates, the force $\bm{F}_{2 \rightarrow 1}$ then simply writes: 

\begin{equation}
    \bm{F}_{2 \rightarrow 1} = \frac{m_1^* m_2^* g^2}{4 \pi \gamma} \frac{\cosh \tau_1 - \cos \sigma_1}{c_2}\bm{e}_{\tau_1}
    \label{Force_bipolar}
\end{equation}

\noindent with $c_2 \simeq (L^2 - r_2^2)/(2r_2)$, set by the geometry of the problem. The force therefore depends on the positions of both particles 1 and 2 in the film. 

In our experiments however, particle 1 remains close to the center of the film, so that $r_1 \ll L$. In this limit, the interaction force $F_{2 \rightarrow 1}$ can be expressed as a function of $d$ and $r_2$ only (see Supplementary Materials), and its amplitude reduces to: 

\begin{equation}
    F_{2 \rightarrow 1} = \frac{m_1^* m_2^* g^2}{4 \pi \gamma d} \left [1-\left (\frac{r_2}{L}\right)^2\right].
    \label{Force_cartesian}
\end{equation}

\noindent This prediction is confirmed experimentally: in Fig.~\ref{Figure2}c, $F_{2 \rightarrow 1}$ deviates from the $1/d$ scaling for $r_2/L > 0.5$. By contrast, the force exerted on particle 2, $F_{1 \rightarrow 2}$ (Fig.~\ref{Figure2}d) follows an $1/d$ law, consistent with the condition $r_1/L < 0.4$ at all times. This asymmetry reflects the breaking of translational invariance in a finite soap film, in contrast with what happens at the surface of a bath. Experimentally, the force ratio $F_{1 \rightarrow 2}/F_{2 \rightarrow 1}$ reaches values up to 1.5. 

\smallskip

To be more quantitative, the exact theoretical prediction (Eq.~\ref{Force_bipolar}, black line) is compared to the experimental measurement of the force in Figures \ref{Figure2} c and d. The theoretical force exhibits small loops, reflecting configurations where the inter-particle distance $d$ is identical, but the radial positions $r_1$ and $r_2$ differ. These loops are not observed experimentally due the uncertainty of the measurements. Nevertheless, the theoretical prediction matches remarkably well the experimental data without adjustable parameter, for both $F_{2 \rightarrow 1}$ and $F_{1 \rightarrow 2}$. This agreement holds for all experiments we performed (see Supplementary Figures 2 and 3), provided that the particles acceleration varies sufficiently smoothly to limit the numerical noise arising from time differentiation.

Another notable feature of Eq. \ref{Force_bipolar} is that  $\bm{F}_{2 \to 1}$ is not aligned with the inter-particle direction $\bm{r}_2 - \bm{r}_1$, as would be expected from a classical interaction force. It is instead oriented along $\bm{e}_{\tau_1}$ (see Figure \ref{Figure3}a). The angle between $\bm{e}_{\tau_1}$ and $\bm{r}_2 - \bm{r}_1$ depends on the positions of the two particles within the film. In our experiments, this angle typically varies between by 0° (when the particles and the film center are aligned) and 15°. Although significant, this deviation remains too small to be clearly observable experimentally (see Supplementary Figure 4). 

To allow a better visualization of how $F_{2 \rightarrow 1}$ and $F_{1 \rightarrow 2}$ vary as the particles move into the film, we show in Figure \ref{Figure3}b-c two representative configurations. The force $F_{2 \rightarrow 1}$ is shown in orange, and $F_{1 \rightarrow 2}$ in green. The amplitude and the direction of two the forces is calculated theoretically using Eq. \ref{Force_bipolar}. Figure \ref{Figure3}b is a situation encountered experimentally, in which $F_{2 \rightarrow 1}$ and $F_{1 \rightarrow 2}$ have very different magnitudes. In this case, the force direction is close to $\bm{r}_2 - \bm{r}_1$, but not strictly equal to it. Figure \ref{Figure3}c shows a symmetric configuration ($r_1 = r_2$), not encountered experimentally, where the angle between the force and $\bm{r}_2 - \bm{r}_1$ reaches 25°.

\section{Infinite soap films}

The interaction force recovers its classical attributes—namely, an orientation along the interparticle direction and a dependence solely on the interparticle distance $d$ in the situation of an infinite film, \textit{i.e.}, which here happens when both particles are located close to the center of the film: $r_1 \ll L$ and $r_2 \ll L$. In this regime, the force writes: 

\begin{equation}
\bm{F}_{2 \rightarrow 1} = - \bm{F}_{1 \rightarrow 2} = \frac{m_1^* m_2^* g^2}{4 \pi \gamma d^2} (\bm{r}_2 - \bm{r}_1),
\label{force_symmetric}
\end{equation}

\noindent a simplified expression also predicted by \cite{martischang2025}. 

To check this model, we systematically varied the radius $R$ (by a factor 3) and the density $\rho$ of the particles (by a factor 4) and measured the interaction force by two different methods. First, the force is measured dynamically (as before), but focusing on orbits where $r_1/L$ and $r_2/L$ are both smaller than $0.35$. In addition, the interaction force is also measured with a \textit{static} method, by considering the equilibrium of paramagnetic particles when placed in a vertical magnetic field \cite{delens2023}. This technique is more precise than the dynamic method, but with a measurement range limited to a few millimeters.  

\smallskip

For the static measurement of the interaction force, we use two soft iron beads (AISI 5100, with density $\rho$ = 7810 kg/m$^3$), placed at the center of two Helmholtz coils, producing a uniform vertical magnetic field $\bm{B} = B\bm{e_z}$. The magnetic field induces a vertical magnetization of the particles, with a magnetic moment $\mu_i = \chi \Omega_i \vec{B}/\mu_0$ for particle $i$ (i = \{1,2\}); noting $\chi = 3$ the effective susceptibility of the particles, $\Omega_i$ the volume of the particle and $\mu_0$ the vacuum permeability. This induces a repulsive dipole-dipole force in the plane of the soap film, of amplitude $F_{\rm mag} = (3 \chi^2 \Omega_1 \Omega_2 B^2)/(4 \pi \mu_0 d^4)$ \cite{lagubeau2016, delens2023}. In presence of the magnetic field, the two particles thus stabilise in the film at a distance $d$ from each other, on each side of the film center. The equilibrium imposes that $F_{\rm mag}$ balances the attractive interaction force -- with a small contribution of the force $F_{\rm 0}$ which can be asymmetric if the particles have different masses (see Supplementary Figure 4 for the details). Experimentally, the magnetic field $B$ is varied between 10 and 35 mT, and the equilibrium distance $d$ is measured once the particles are stabilized at their equilibrium position, typically 10 seconds after being deposited in the film. 
Assuming the symmetry of the interaction force, which is valid for $r_1/L$ and $r_2/L <$ 0.2, the measurement of $d$ gives $F_{2 \rightarrow 1} = F_{\rm mag}(d) - F_0(r_1)$.

\begin{figure}
    \centering
    \includegraphics[width=1\linewidth]{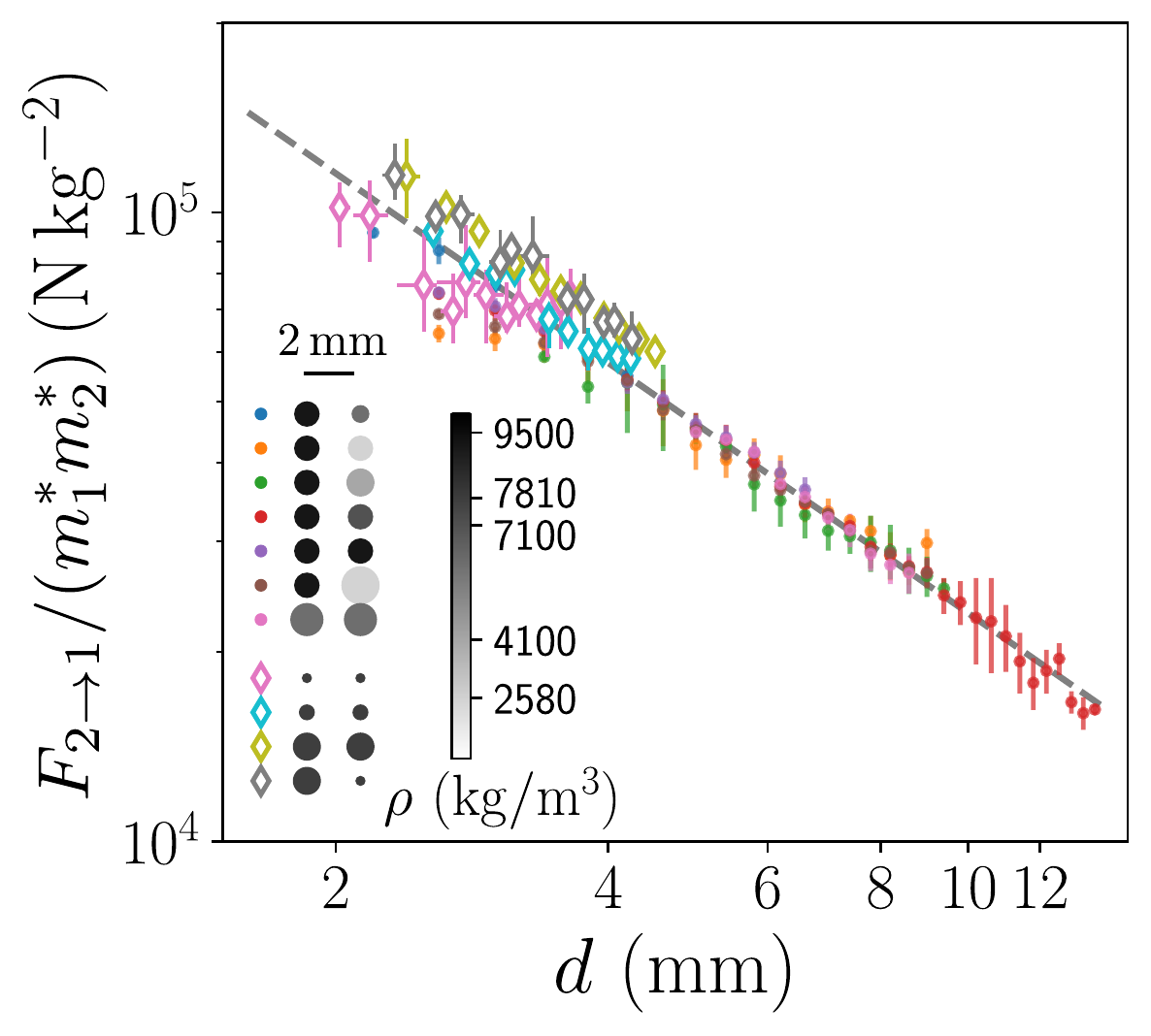}
    \caption{Interaction force $F_{2 \rightarrow 1}$ for $r_1 \ll L$ and $r_2 \ll L$. Here $F_{2 \rightarrow 1}$ is divided by the product of the masses of the two particles $m_1^* m_2^*$, and plotted as a function of the interparticle distance $d$. The colored dots correspond to the dynamic measurements (with error bars corresponding to the standard deviation of the force for one experiment and for $d$ varied by 400 µm). The white diamonds correspond to magnetic measurements (with error bars corresponding to 3 different measurements). The legend indicates the radius of the solid particles (varied by a factor 4) and the gray scale gives their density (varied also by a factor 4). All data collapse on a single curve of slope -1 in log-log. The dotted line shows the theoretical prediction (Equation \ref{force_symmetric})}
    \label{Figure4}
\end{figure}

\bigskip

\noindent Figure \ref{Figure4} compiles all measurements of the interaction force $F_{2 \rightarrow 1}$, both obtained dynamically (filled circles) and using the magnetic actuation of the particles (empty diamonds). For each experiment, the legend gives the radius $R$ of the particles (left: particle 1 and right: particle 2) and their density $\rho$ (gray scale). Note that the masses of the moving objects differ from the masses of the central spheres due to the additional mass of the meniscus, which is measured independently for each data point. Here, the force $F_{2 \rightarrow 1}$ is divided by product of the masses $m_1^* m_2^*$ of the two moving objects. Using this representation, all the experiments -- both dynamic and magnetic data, for particle radii varying by a factor 4 and density by a factor 3 collapse on a single curve, function of the interparticle distance $d$ only. The dotted gray line is the theoretical prediction (Eq. \ref{force_symmetric}) which matches all experimental data without adjustable parameter.

\section{\label{Conclusion}Conclusion}

Despite being also mediated by a deformation of the liquid-air interface, the attraction force between two particles in a soap film fundamentally differs from what is classically seen on a bath. The difference happens at two levels: first, the force has an extremely long range (up to the size of the film), which, combined to low friction induces striking particle dynamics. The two beads join in a complex dance where they orbit each other for typically 10 seconds. Second, and even more unusual, the attraction force looses the general attributes of an interaction force: it is not a function of the inter-particle distance $d$ only, but depends on the respective positions of the two particles in the film, so that $F_{2 \rightarrow 1} \neq F_{1 \rightarrow 2}$. We evidence experimentally this phenomenon: in situations where one particle is close to the center of the film and the other is close to the frame, the imbalance $F_{2 \rightarrow 1}/F_{1 \rightarrow 2}$ can reach a factor 1.5. Our theoretical model predicts exactly this phenomenon, and also evidences an asymmetry in the direction of the force, which is not strictly oriented along the interparticle direction. 
The combination, in a soap film, of a very-long range attraction force and low friction opens new possibilities for controlled particle self-assembly and the engineering of 2D materials.


\bibliographystyle{unsrt}
\bibliography{biblio.bib}

\end{document}